\documentclass[letterpaper]{article} 
\usepackage{aaai2026}  
\usepackage{times}  
\usepackage{helvet}  
\usepackage{courier}  
\usepackage[hyphens]{url}  
\usepackage{graphicx} 
\usepackage{natbib}  
\usepackage{caption} 
\usepackage{algorithm}
\usepackage{natbib}
\usepackage{xcolor}
\usepackage{listings}
\definecolor{codebg}{gray}{0.95}  

\floatstyle{ruled}
\newfloat{listing}{tb}{lst}{}
\floatname{listing}{Listing}
\usepackage{newfloat}
\usepackage{listings}
\usepackage{booktabs}
\usepackage{tikz}
\usepackage{caption}
\usepackage[most]{tcolorbox}
\usepackage{multirow}
\usepackage{array}

\usepackage{xurl}
\DeclareCaptionStyle{ruled}{labelfont=normalfont,labelsep=colon,strut=off} 
\floatstyle{ruled}
\newfloat{listing}{tb}{lst}{}
\floatname{listing}{Listing}
\title{LSDTs: LLM-Augmented Semantic Digital Twins for Adaptive\\ Knowledge-Intensive Infrastructure  Planning }

\author{
    Naiyi Li\textsuperscript{\rm 1}\thanks{Co-lead authors.},
    Zihui Ma\textsuperscript{\rm 1,2}\footnotemark[1],
    Runlong Yu\textsuperscript{\rm 3}\thanks{Co-corresponding authors.},
    Lingyao Li\textsuperscript{\rm 4}\footnotemark[2]
}
\affiliations{
    \textsuperscript{\rm 1}Department of Civil \& Environmental Engineering, University of Maryland, College Park\\
    \textsuperscript{\rm 2}Center for Urban Science and Progress, New York University\\
    \textsuperscript{\rm 3}Department of Computer Science, University of Alabama\\
    \textsuperscript{\rm 4}School of Information, University of South Florida\\
    nli5211@umd.edu, zihuima@nyu.edu, ryu5@ua.edu, lingyaol@usf.edu
}
\usepackage{bibentry}

\DeclareUnicodeCharacter{202F}{\,}
\begin{document}

\maketitle

\begin{abstract}


Digital Twins (DTs) offer powerful tools for managing complex infrastructure systems, but their effectiveness is often limited by challenges in integrating unstructured knowledge. Recent advances in Large Language Models (LLMs) bring new potential to address this gap, with strong abilities in extracting and organizing diverse textual information. We therefore propose LSDTs (LLM-Augmented Semantic Digital Twins), a framework that helps LLMs extract planning knowledge from unstructured documents like environmental regulations and technical guidelines, and organize it into a formal ontology. This ontology forms a semantic layer that powers a digital twin—a virtual model of the physical system—allowing it to simulate realistic, regulation-aware planning scenarios. We evaluate LSDTs through a case study of offshore wind farm planning in Maryland, including its application during Hurricane Sandy. Results demonstrate that LSDTs support interpretable, regulation-aware layout optimization, enable high-fidelity simulation, and enhance adaptability in infrastructure planning. This work shows the potential of combining generative AI with digital twins to support complex, knowledge-driven planning tasks.


\end{abstract}

\section{Introduction} 

Large-scale infrastructure projects (e.g., energy systems, transportation networks) often face significant challenges in project planning \cite{flyvbjerg2007policy, flyvbjerg2009survival}. A key factor underlying management failures is the lack of effective knowledge integration \cite{huang2022information, d2020bim}, as critical project information often remains fragmented across unstructured documents---such as contracts specifying environmental impact assessments, regulatory codes mandating safety standards. Conventional project management frameworks (e.g., PMBOK~\cite{project2021guide}, PRINCE2~\cite{bentley2012prince2}) are not designed to process or reason over such content. Although Digital Twins (DTs) have transformed infrastructure planning through real-time monitoring and predictive modeling~\cite{broo2022design, broo2023digital, deng2025integrating}, they often depend on structured inputs from Internet of Things (IoT) models~\cite{minerva2020digital, kaur2019convergence, zhao2021iot}. As DT technology advances, there is an urgent need to enable DTs to incorporate more advanced language processing capabilities for information retrieval and integration.


To address these challenges, many DT studies have leveraged sematic web technologies \cite{jiang2023intelligent, jagatheesaperumal2023semantic} for structured information extraction. In addition, researchers have applied natural language processing (NLP) tools such as rule-based \cite{su2022research} or BERT-based models \cite{jiang5344488automated} for information extraction. However, challenges still remain. Current methods often struggle to transform unstructured, large-scale text into structured, machine-readable knowledge. They also have limited ability to integrate textual content with spatial representations, reason across documents, or adapt dynamically as real-time planning information become available in DTs.


Recent advances in Large Language Models (LLMs) offer a promising path forward \cite{wang2024twin, li2023chattwin, chen2025integrating}. Models like GPT-4 and Gemini demonstrate strong capabilities in understanding documents, synthesizing information across sources, and reasoning about complex relationships \cite{fan2024nphardeval, dagdelen2024structured}. LLMs can extract key entities (e.g., stakeholders, deadlines, and constraints), identify dependencies between requirements and regulations, and reason semantically across documents \cite{sun2025docs2kg, zhou2025named}, with minimal need for manual rule creation. These strengths enable a more scalable and context-aware approach to managing complex project information.

Therefore, we propose \textbf{LLM-Augmented Semantic Digital Twins (LSDTs)}—an integrated architecture that combines the language understanding of LLMs with the simulation and visualization strengths of DTs. Our work makes two main contributions. First, we develop \textbf{an LLM-driven pipeline for knowledge structuring}, where LLMs extract regulatory constraints and planning knowledge from unstructured documents into a structured semantic graph that unifies diverse regulatory, spatial, and technical requirements. Second, we design \textbf{a system workflow for DTs} that links LLM-derived knowledge with real-time simulation and scenario analysis. Our developed system provides important capabilities for automatically interpreting planning documents, structuring information, and supporting adaptive, regulation-aware decision-making in infrastructure projects.



\section{Related Work}


\subsection{Infrastructure Planning and Digital Twins}

Infrastructure planning for large-scale projects relies on structured semantic models, such as IFC-enabled building information modeling (BIM), which captures detailed building-level geometry and semantics, and CityGML, which extends these representations to the urban context \cite{lam2024smartcity}. Despite these advances, incorporating critical but unstructured information, such as environmental regulations, stakeholder requirements, and policy guidelines, remains challenging, especially during early planning stages. Existing ontology-based systems can extract specific regulatory clauses from textual sources, but they often require extensive manual rule definitions and lack adaptability \cite{peng2023acc}.

In parallel, digital twin (DT) technologies have matured around operational applications, such as sensor-driven monitoring and predictive maintenance~\citet{wahab2024pmdt}, while their use in the early-stage planning are largely experimental and conceptual \cite{mousavi2024review}. Recent advances, such as asset administration shells (AAS) integrated with information containers for linked document delivery (ICDD), demonstrate how modular and interoperable DTs can support project management \cite{zhang2025aas}. Similarly, microservice-based DTs pilots in domains such as smart grids and climate-resilient infrastructure highlight the potential of loosely coupled and adaptable architectures \cite{mchirgui2024smartgrid}. However, current DTs rarely integrate unstructured textual data seamlessly or accommodate dynamic, stakeholder-specific constraints in planning contexts. This limitation is particularly pronounced in wind farm planning \cite{ambarita2024industrial}, where planners must reconcile spatial models, environment assessments, and policy documents.


\subsection{LLM Application in Infrastructure Planning}

LLMs have demonstrated robust capabilities in entity extraction, relation identification, cross-document reasoning, and multimodal comprehension. These models efficiently transform heterogeneous inputs, including PDFs, HTML pages, and plain text, into structured knowledge without extensive manual engineering. Early built-environment applications exemplify this potential: ~\citet{holtgen2025semantic} employ LLMs to automatically convert relational construction data into Resource Description Framework (RDF) semantic graphs, significantly streamlining ontology mapping. ~\citet{madireddy2025code} leverage LLM pipelines to translate building codes into machine-readable rules for BIM compliance checking. ~\citet{du2024text2bim} introduce Text2BIM, a multiagentframework that generates editable 3D building models from natural language descriptions. Moving beyond single-asset contexts, ~\citet{gurcan2025socialDT} embeds LLM agents within a Social Digital Twinner platform, enabling natural language querying and steering of agent-based simulations, thus highlighting a novel interaction paradigm for policy-oriented digital twins. A recent survey by ~\citet{yang2025review} organizes these emerging efforts into a descriptive–predictive–prescriptive taxonomy, identifying critical challenges including data grounding, continuous model updating, and provenance tracking.

Despite these promising advances, LLM applications in infrastructure planning remain largely limited to isolated linguistic tasks, such as classification, summarization, and question-answering, without deep integration into simulation engines or decision-support workflows. Current prototypes seldom inject LLM-extracted facts into persistent semantic graphs, lack mechanisms for tracking evolving planning constraints, and rarely link outputs directly to scenario analytics or optimization pipelines. 


\section{Methodology}

\begin{figure*}[t]
    \centering
    \includegraphics[width=0.98\textwidth]{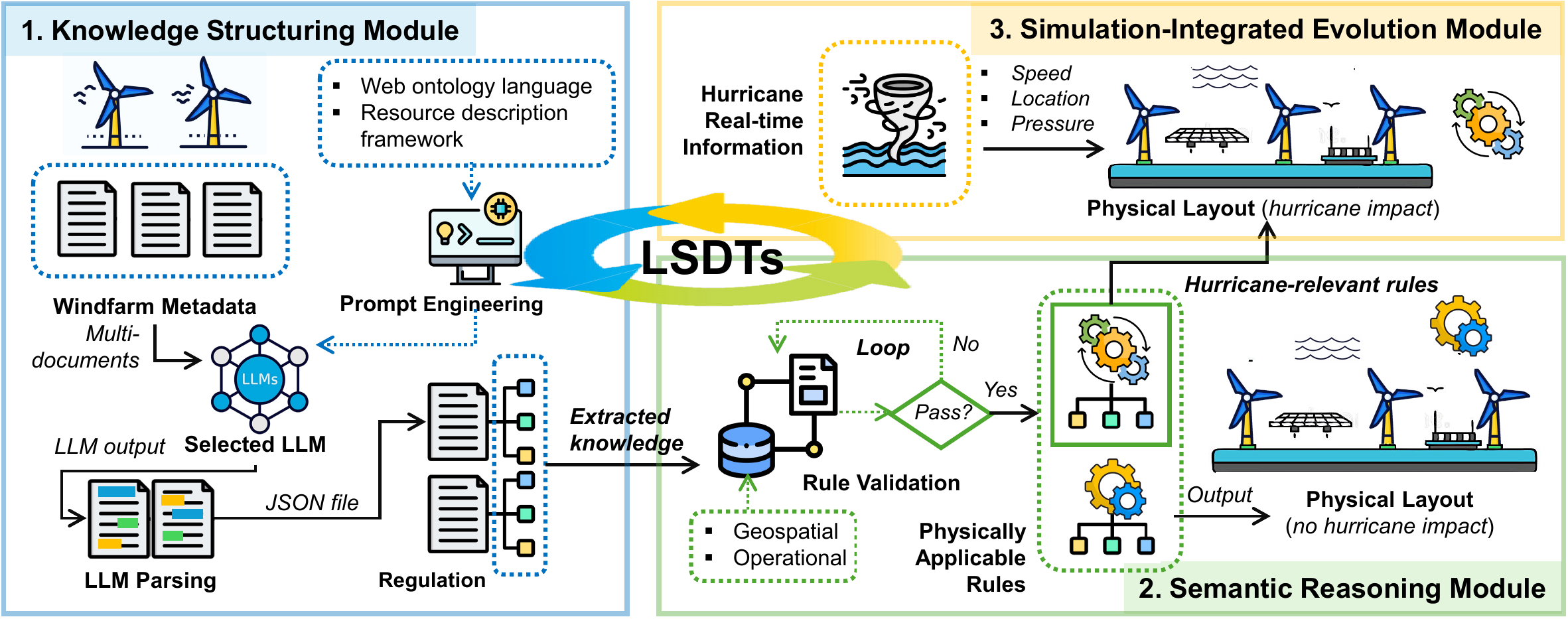}
    \caption{An illustrative framework design of LSDTs.}
    \label{fig: framework}
\end{figure*}

This section introduces our conceptual framework, LLM-Augmented Semantic Digital Twins (LSDTs), a knowledge-guided planning architecture that integrates LLM-based knowledge extraction, ontology-driven reasoning, and environmental simulation within a unified semantic substrate (Figure~\ref{fig: framework}). LSDTs prioritize structured knowledge construction and guided decision support. Rather than treating compliance, simulation, and optimization as disconnected stages, the framework operates as a closed-loop workflow in which each component iteratively updates, queries, and enriches a shared semantic graph. At the center of this loop is a domain-specific ontology that shapes the structure of extracted content, constrains simulation behaviors, and mediates planning decisions. This knowledge-centered design ensures semantic alignment and interpretability across all modules, enabling explainable and regulation-aware infrastructure planning without resorting to black-box automation.

\subsection{Semantic Foundations}

At the core of the LSDT framework is a knowledge structure $G = \{(s, p, o)\}$ composed of RDF triples that instantiate a domain ontology $\mathcal{O}$. The ontology defines key concepts, relationships, and geospatial constraints relevant to infrastructure planning—for example, class hierarchies (e.g., \textit{WindFarm} $\sqsubseteq$ \textit{Infrastructure}), property domains and ranges (e.g., \textit{intersects}: \textit{Geometry} $\times$ \textit{Geometry}), and built-in geospatial functions wrapped around the Java Topology Suite~\footnote{https://locationtech.github.io/jts/}. These axioms provide the shared vocabulary across modules and support schema validation, logical inference, and spatial reasoning. Given a corpus of unstructured planning documents $\mathcal{D}$, we define a language model function:
\begin{equation}
L: \mathcal{D} \rightarrow \mathcal{I} = \mathcal{I}_A \cup \mathcal{I}_C
\end{equation}
that extracts two kinds of semantically typed information: 
\begin{itemize}
    \item \textbf{Attribute snippets} $\mathcal{I}_A$: entity-level properties (e.g., turbine spacing, setback distances);
    \item \textbf{Constraint snippets} $\mathcal{I}_C$: regulation-like statements describing permitted or prohibited configurations.
\end{itemize}

These outputs are not free-form: they are guided by ontology-aware prompts and post-processed to ensure consistency with $\mathcal{O}$. Attribute-level statements are converted to RDF triples using a schema-aligned instantiation function $F_{\text{inst}}: \mathcal{I}_A \rightarrow \mathcal{T}^+$, producing an initial graph $G_0$. Constraint expressions are formalized as executable logical rules via $F_{\text{rule}}: \mathcal{I}_C \rightarrow \mathcal{R}$, where each rule $r \in \mathcal{R}$ takes the form:
\begin{equation}
\text{Body}(r) \Rightarrow \text{Head}(r)
\end{equation}
with Body$(r)$ representing conjunctions of triple patterns and spatial predicates, and Head$(r)$ asserting derived facts (e.g., \textit{:hasConflict true}) into the graph.

Given a candidate design $\Delta G$ (e.g.,a proposed wind farm layout), we update the graph $G = G_0 \cup \Delta G$ and apply a forward-chaining reasoner to evaluate the rule set:
\begin{equation}
G^\ast = \mathrm{Reason}(G, \mathcal{R})
\end{equation}
Violations or inferences are explicitly materialized as new triples, enabling transparent, queryable validation.

To incorporate dynamic environmental inputs, such as wind, weather, or hazards, we define a simulation function $\Psi$ that injects time-varying exogenous data $S_t$ into the graph:
\begin{equation}
G_t = \Psi(G_{t-1}, S_t)
\end{equation}
This produces a sequence $\{G_t\}_{t=0}^T$ representing the evolving semantic state of the infrastructure system under changing conditions. At each time step, $\mathcal{R}$ is reevaluated on $G_t$ to assess regulatory compliance under evolving conditions.

Finally, an external optimization function $\mathcal{O}pt$ queries the graph $G^\ast$ for constraint violations or performance metrics, and proposes revised candidate designs $\delta G$, closing the loop. This architecture supports iterative refinement, where knowledge extraction, reasoning, simulation, and design co-evolve within a shared semantic substrate.

\subsection{Modular Implementation}

Building upon the core architecture of LSDTs, we now turn to its system-level implementation, which enables real-time planning and adaptive simulation based on the semantic foundation. While the previous section introduced the formalism underlying the framework, this section details how its components are integrated into a three-module workflow in DTs. Specifically, we demonstrate how LLMs enable the construction of a structured semantic graph, how LLM-derived planning constraints guide scenario simulations, and how environmental inputs are dynamically injected into the semantic substrate to support time-sensitive and regulation-aware decision-making.

\noindent\textbf{Module 1: Knowledge Structuring.}
This module uses knowledge-guided prompt templates to extract semantic information from unstructured planning documents. Leveraging LLMs, it produces attribute-level facts $\mathcal{I}_A$ and constraint-level expressions $\mathcal{I}_C$, each anchored to the predefined semantic schema. 
The outputs are post-processed to normalize formats, resolve units, and ensure consistency with the domain ontology. Attribute facts are inserted into the initial RDF graph $G_0$ via type-safe instantiation, while constraint expressions are compiled into executable rules that form the rule set $\mathcal{R}$. This automated pipeline enables reliable knowledge acquisition with minimal human supervision.

\noindent\textbf{Module 2: Semantic Reasoning.}
Once knowledge is structured, it should be used to assess whether candidate designs meet all applicable constraints continuously. Traditional workflows defer compliance checks to post-design phases, often requiring costly revisions. This module embeds compliance logic directly into the semantic graph by executing forward chain inference on $G = G_0 \cup \delta G$ using a set of rules $\mathcal{R}$. Potential violations are materialized as triples RDF (for example, asserting \textit{:hasConflict true} on a non-compliant entity) within the updated graph $G^\ast$, allowing real-time validation, transparent explanations, and machine-interpretable feedback. This tight coupling between design and reasoning supports iterative layout refinement and speeds up convergence toward regulation-compliant configurations.

\noindent\textbf{Module 3: Simulation-Integrated Evolution.}
Infrastructure should also withstand uncertain and time-varying environmental factors, such as extreme weather events. However, resilience assessment is often disconnected from design workflows. This module introduces a simulation function $\Psi$ that injects exogenous, time-indexed inputs $S_t$ (e.g., storm trajectories, sensor readings) into the graph. Simulation results, such as turbine shutdowns or damage estimates, are expressed using the same ontology $\mathcal{O}$, allowing downstream rules in $\mathcal{R}$ to be re-applied at each timestep. This integration supports compliance-aware adaptation under dynamic conditions, enabling planners to evaluate both feasibility and resilience in a unified reasoning loop.


\subsection{Instantiation: Offshore Wind Farm Planning}

To demonstrate the generality of LSDTs, we instantiate the framework for offshore wind farm planning, a representative infrastructure task with complex spatial regulations and environmental dependencies. In this case, the ontology $\mathcal{O}_{\text{INFRASTRUCTURE}}$ includes domain-specific classes such as WindFarm, Turbine, and Cable, with geometry-typed properties defined according to geospatial standards. LLMs including GPT-4o-mini~\cite{openai2024gpt4o}, Claude Haiku 3.5~\cite{anthropic2024claude35sonnet}, Gemini-1.5-Pro~\cite{google2024gemini15}, and DeepSeek \citep{deepseek2024v2} are applied to regulatory documents from various documentations, yielding structured RDF outputs $\mathcal{I}_A$ and constraint expressions $\mathcal{I}_C$. After schema validation, we obtain an initial graph $G_0$ and a rule set $\mathcal{R}$ encoding constraints such as minimum turbine spacing and exclusion buffers around protected zones.

Candidate layouts are encoded as RDF triples using Well-Known Text (WKT) literals and added to the base graph as $\delta G$. We employ the Jena rule engine~\footnote{https://jena.apache.org/}, extended with geospatial built-ins, to apply the previously defined rule set $\mathcal{R}$. This module operationalizes the reasoning process described earlier, enabling automated detection of spatial conflicts (e.g., overlap with protected zones) within the semantic graph.

To incorporate environmental stressors, we simulate storm conditions using historical data from NOAA’s HURDAT2 dataset. For each time step $t$, we compute the wind field using the Holland profile model and apply derived wind speeds for detailed calculations) to turbine locations. Each turbine is assigned a status (e.g., \textit{:Shutdown}) based on cutout thresholds, and the simulation function $\Psi$ inserts these into the evolving graph state $G_t$. Constraint rules are re-applied over this state, enabling the planner to identify vulnerable configurations.

An external solver $\mathcal{O}pt$ queries the graph $G_t$ for violations and performance metrics (e.g., power yield), and proposes a revised layout. This optimization loop iterates until a design satisfies all constraints or converges to a fixed point.

\section{Experimental Settings}
Our experimental setup includes the following components:
\paragraph{Regulatory Documents:}
We compile a diverse corpus of publicly available planning documents, including Records of Decision (RODs) for existing U.S. wind leases, Environmental Impact Statements (EISs), and technical specifications. To ensure comprehensive coverage, we further incorporate key legislative documents such as the Outer Continental Shelf Lands Act (OCSLA) \cite{ocsla}, the National Environmental Policy Act (NEPA) \cite{nepa}, the Energy Policy Act of 2005 \cite{epact2005}, and the Marine Mammal Protection Act (MMPA) \cite{mmpa}. These sources provide a mixture of spatial, ecological, and procedural constraints to test LLMs in extracting planning knowledge. 

\begin{table*}[htbp]
\centering
\footnotesize  
\setlength{\tabcolsep}{5pt} 

\begin{tabular}{c cc cc cc cc}
\toprule
\textbf{Document No.} 
& \multicolumn{2}{c}{\textbf{Claude 3.5 Haiku}} 
& \multicolumn{2}{c}{\textbf{Gemini 1.5 Pro}} 
& \multicolumn{2}{c}{\textbf{GPT-4o-mini}} 
& \multicolumn{2}{c}{\textbf{DeepSeek-chat}} \\
\cmidrule(r){2-3} \cmidrule(r){4-5} \cmidrule(r){6-7} \cmidrule(r){8-9}
& \textbf{Acc.} & \textbf{No. of Reg.}
& \textbf{Acc.} & \textbf{No. of Reg.}
& \textbf{Acc.} & \textbf{No. of Reg.}
& \textbf{Acc.} & \textbf{No. of Reg.} \\
\midrule
17 & 0.800 & 15 & 0.692 & 13 & 1.000 & 8  & 0.692 & 13 \\
20 & 0.875 & 16 & 0.950 & 20 & 0.800 & 5  & 0.944 & 18 \\
21 & 0.778 & 18 & 0.870 & 23 & 0.900 & 10 & 0.895 & 19 \\
27 & 0.778 & 15 & 0.913 & 23 & 0.750 & 8  & 0.625 & 8  \\
\bottomrule
\end{tabular}
\caption{Model performance evaluations.}
\label{tab:model-accuracy-rules}
\end{table*}

\paragraph{Layout Optimization Setup:}

\begin{figure*}[t]
    \centering
    \includegraphics[width=\textwidth, height=0.9\textheight, keepaspectratio]{
    	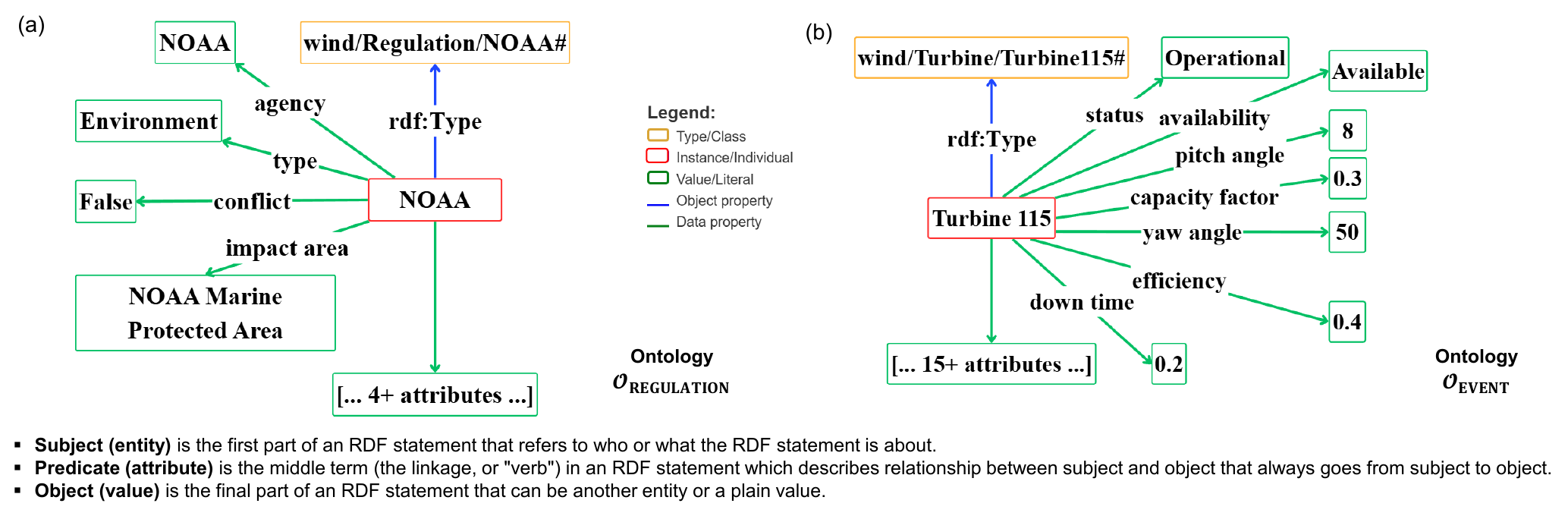}
    \caption{Example knowledge structures for (a) Regulation and (b) Event domains.}
    \label{fig:ontology}
\end{figure*}

We present a pilot study focused on the Maryland Offshore Wind Lease Area (OCS-A 0490), an approved lease zone along the U.S. Atlantic coast (Figure~\ref{fig:layout}(b)). 
This site is selected due to its regulatory complexity and the fact that construction has not yet begun, making it a timely and realistic testbed for early-stage planning analysis. To provide a broader spatial context, Figure~\ref{fig:layout}(a) illustrates the national distribution of offshore wind development zones, including marine protected areas, military exclusion zones, commercial shipping corridors, bathymetric features, and designated lease boundaries. These layers, sourced from the MarineCadastre.gov GIS database \cite{boem2024marinecadastre}, serve as the geospatial foundation for our modeling pipeline. 

To encode and reason over planning constraints, we adopt Semantic Web technologies. Regulatory requirements extracted by LLMs are formalized into RDF triples, using OWL ontologies \cite{w3c_owl2_2012} to define spatial concepts, turbine properties, and regulatory constraints. Constraint enforcement is implemented using the Jena rule engine, which enables logical inference over turbine layouts. 
These structured constraints guide schematic layout generation for 121 turbines arranged in 13 rows, as specified by the Maryland lease design requirements. We use the TopFarm \cite{topfarm2013} Python package to optimize turbine coordinates, modeling wake interactions via the PyWake library~\footnote{https://topfarm.pages.windenergy.dtu.dk/PyWake/} and the Bastankhah-Gaussian model. The turbine model follows the IEA 15-MW reference specification (240\,m rotor diameter, 150\,m hub height) \cite{iea15mw2020}. Wind conditions are represented using a 24-sector wind rose, each modeled with a Weibull distribution ($k=2.5$, $A=8.0$). The primary objective is to maximize the Annual Energy Production (AEP), a key metric of economic viability. Optimization is subject to two hard constraints: (1) a minimum spacing of five rotor diameters (1,200\,m) to mitigate wake losses, and (2) strict adherence to polygonal lease boundaries defined via WKT geometry. We employ Stochastic Gradient Descent (SGD) over 400 iterations to arrive at the final layout.

\paragraph{Hurricane Simulation Setup:}
We assess hurricane simulations implemented in Python. Hurricane Sandy (2012) serves as a high-impact test case, selected for its status as one of the most destructive storms in recent Maryland history. It made landfall near Brigantine, New Jersey, on October~29, with sustained winds of approximately 80~mph (129~km/h), resulting in extensive flooding and infrastructure disruptions across the Chesapeake Bay region \cite{zimmerman2014planning, njdep_sandy10, NWS_HurricaneSandy}. Figure \ref{fig: results}(a) illustrates its trajectory. 
Key simulation parameters include a hub height of 150~m, a cut-out wind speed of 25~m/s, a Holland parameter of $B = 1.5$, and a radius of maximum winds of 50~km.
Turbine response is governed by a shutdown protocol that activates when two conditions are met: (1) the storm enters a defined proximity threshold, and (2) hub-height wind speeds exceed the turbine’s operational cut-out limit. The semantic control model monitors three key turbine-level actions: pitch angle (e.g., blade feathering), yaw angle (e.g., alignment to wind), and operational status (e.g., shutdown). All turbine actions are logged in RDF format, generating a machine-readable operational states. Additional outputs include a time-stamped CSV of wind metrics, turbine states, and storm positions.

\section{Experimental Results}

Building on our proposed LSDTs framework and experimental design, this section presents results across three key modules of the pipeline.

\subsection{Knowledge Structuring Performance}
We structure our knowledge base using three core domains—infrastructure ($\mathcal{O}_{\text{INFRASTRUCTURE}}$), event ($\mathcal{O}_{\text{EVENT}}$), and regulation ($\mathcal{O}_{\text{REGULATION}}$). Each domain is supported by a dedicated ontology and follows the same LLM-driven pipeline for schema-compliant knowledge extraction. In this section, we focus on evaluating the performance of our knowledge structuring module in the regulatory domain ($\mathcal{O}_{\text{REGULATION}}$). Figure~\ref{fig:ontology}(a) illustrates a representative RDF graph from this domain. It encodes a constraint involving the National Oceanic and Atmospheric Administration (NOAA), with properties such as \textit{:hasRegulationDescription}, \textit{:hasImpactArea}, and \textit{:hasConflict}. The \textit{:hasConflict} property serves as a dynamic compliance flag, updated during semantic reasoning to indicate whether the wind farm violates this rule. The zoomed-in view highlights the \textit{False} value assigned to this property, confirming that the current layout satisfies the constraint. 
\begin{table}[htbp]
\centering
\small
\resizebox{\columnwidth}{!}{%
\begin{tabular}{lrrrr}
\toprule
\textbf{Row} & \textbf{Mean X} & \textbf{Std Dev X} & \textbf{Mean Y} & \textbf{Std Dev Y} \\
\midrule
1  &   68.43 & 61.32 & 136.89 &   87.76 \\
2  &   72.31 & 55.18 &   58.96 & 107.90 \\
3  &   65.14 & 49.18 &   35.24 &   18.17 \\
4  &   41.63 & 38.81 &   49.75 &   32.86 \\
5  &   73.29 & 42.36 &   44.09 &   30.72 \\
6  &   68.01 & 57.71 &   26.10 &   17.50 \\
7  &   66.73 & 66.62 &   28.05 &   21.64 \\
8  &   79.44 & 55.45 &   38.37 &   20.53 \\
9  &   77.18 & 42.92 &   27.16 &   19.89 \\
10 &   67.15 & 63.48 &   33.84 &   12.69 \\
11 &   68.70 & 71.34 &   47.91 &   31.09 \\
12 &   69.40 & 77.09 &   70.12 &   35.50 \\
13 &   82.87 & 63.93 & 109.51 &   39.85 \\
\midrule
\textbf{Overall} & \textbf{67.98} & \textbf{54.83} & \textbf{51.81} & \textbf{54.95} \\
\bottomrule
\end{tabular}%
} 
\caption{Positional deviation (meters).} 
\label{tab:positional-deviation} 
\end{table}
To assess extraction quality, we tested how well LLMs identify and formalize planning constraints from unstructured regulatory documents. First, we selected four representative documents and asked two domain experts to manually annotate constraint regulations to serve as ground truth. The inter-annotator agreement, measured using Krippendorff’s Alpha for detailed calculations), achieved an average score of 0.80, indicating strong consistency and reliable reference data for model comparison.

We then evaluated the performance of our four selected LLMs by comparing their extracted regulations against the human-annotated ground truth. Each model received the same schema-constrained prompt. Prompting strategies such as few-shot examples and chain-of-thought reasoning guide the models to generate RDF-compatible triples. We report two metrics: (1) accuracy, defined as the proportion of correctly matched regulations over the total number of expert-annotated regulations, and (2) regulation count, the total number of regulations each model extracted from the document. A regulation is considered a correct match if its extracted structure semantically aligns with a ground-truth rule, regardless of wording differences. The results are summarized in Table~\ref{tab:model-accuracy-rules}.

Overall, all four models performed well, achieving relatively high accuracy across the evaluated documents. In particular, Gemini 1.5 Pro demonstrated consistently strong performance, combining both high accuracy and broad coverage. It produced the highest number of extracted regulations and achieved top accuracy in three out of four cases. Claude 3.5 Haiku also showed solid results, with balanced extraction quality and high correctness.
We also observed distinct behavioral patterns across the models. For example, GPT-4o-mini reached perfect accuracy (1.000) on one document but extracted fewer regulations overall, suggesting a conservative strategy that favors precision over recall. DeepSeek-chat performed well when regulatory rules are explicitly presented in structured formats but struggled with more implicit or fragmented constraints. This was particularly evident in documents 17 and 22, which focus more on site assessment analysis. In contrast, documents 20 and 21, published by the Bureau of Ocean Energy Management \cite{boemLeases}, contain direct regulatory language and are easier for all models to interpret. Across the four models, a total of 289 unique and semantically validated regulations are extracted and encoded into RDF for downstream reasoning. These results confirm the feasibility of automating schema-compliant knowledge acquisition from heterogeneous regulatory corpora.   

\subsection{LLM-derived Wind Farm Layout}
\begin{figure}[htbp]
    \centering
    \includegraphics[width=0.86\columnwidth]{
    	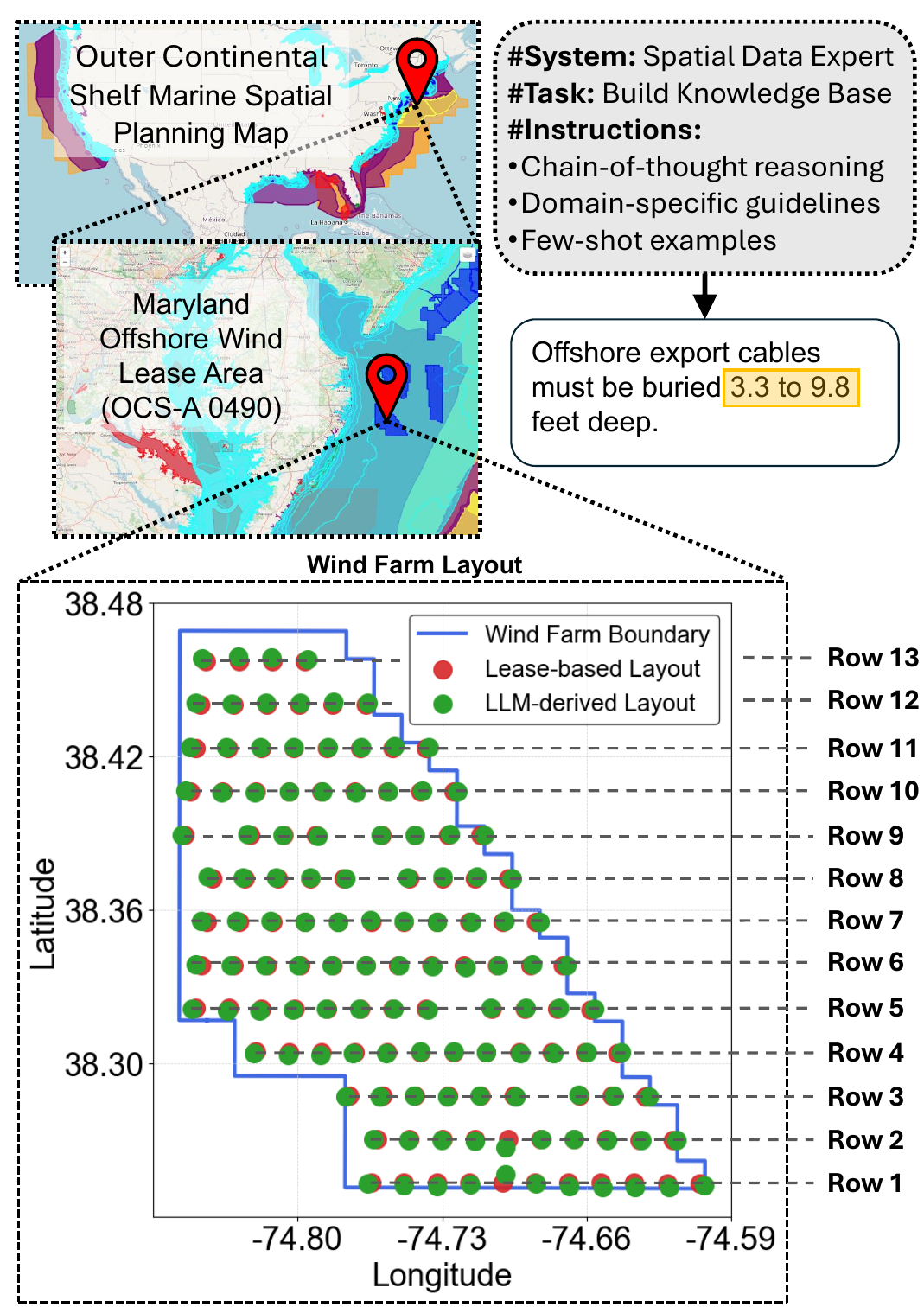}
    \caption{Pilot study of wind farm layout.}
    \label{fig:layout}
\end{figure}

Next, we test whether the structured regulatory knowledge can guide compliant layout generation. Figure~\ref{fig:layout}(c) presents a comparison between the LLM-derived layout (green dots) and the lease-based reference layout (red dots). Both layouts exhibit similar spatial patterns, suggesting that the regulatory graph constructed from text-extracted constraints can effectively guide the generation of realistic turbine configurations. This result demonstrates the feasibility of automating early-stage design workflows using LLMs, even in the absence of direct access to proprietary turbine placement data. To further quantify the spatial alignment between the two layouts, we compute the positional deviation for each row of turbines. Specifically, we calculate the mean and standard deviation of longitude (X) and latitude (Y) differences across corresponding turbines within each row. These statistics capture both the average displacement and the spatial variability between the LLM-optimized and lease-based configurations. Table~\ref{tab:positional-deviation} summarizes the results. 

On average, the LLM-derived turbine locations deviate by approximately 67.98 meters in the X direction and 51.81 meters in the Y direction, with standard deviations of 54.83 meters (X) and 54.95 meters (Y). While rows 1 and 2 exhibit slightly larger deviations due to reduced wake interactions and spacing constraints, the overall magnitude of difference remains within acceptable tolerances for schematic-level planning. These findings show that our LLM-driven framework is capable of producing spatially coherent, regulation-compliant layouts under realistic planning constraints.

\begin{figure*}[htbp]
    \centering
    \includegraphics[width=0.85\textwidth]{
    	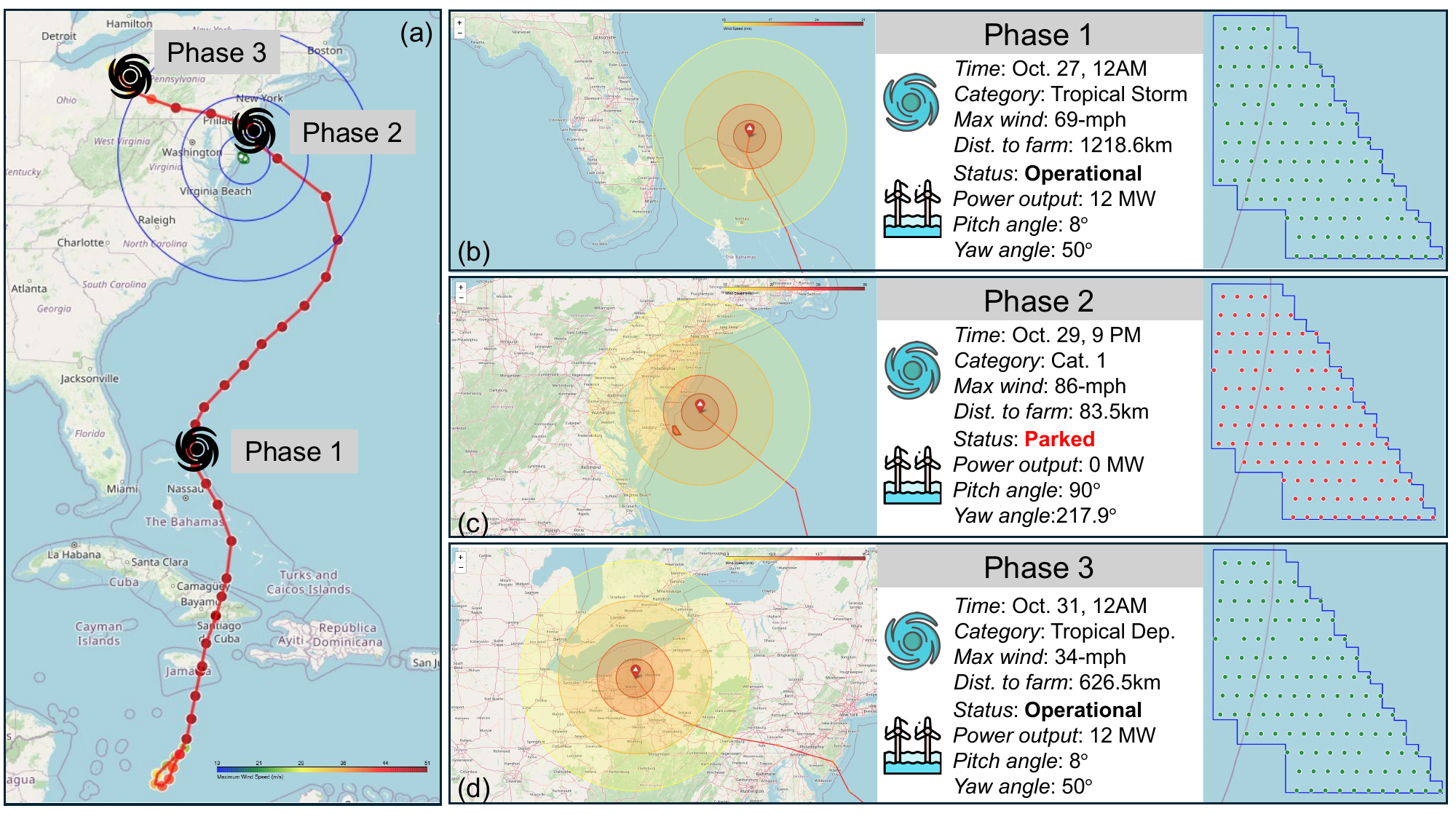}
    \caption{LSDT-driven turbine status changes during 2012 Hurricane Sandy.}
    \label{fig: results}
\end{figure*}

\subsection{LSDTs for Hurricane Sandy Simulation}
Finally, to access system-level adaptability, we simulate the impact of Hurricane Sandy across three phases: (1) \textbf{Phase 1}: Sandy tracks northward toward the Mid-Atlantic coast.  
(2) \textbf{Phase 2}: Tropical-storm-force winds reach Maryland’s Eastern Shore.  
(3) \textbf{Phase 3}: The storm moves inland and offshore wind conditions begin to subside.

During Phase 1 (Figure\ref{fig: results}(b)), all turbines remain operational. Simulated wind speeds are 69\,mph and the distance to the Maryland lease area is approximately 1,219\,km, both of which fall below the intervention threshold. As Sandy advances in Phase~2 (Figure \ref{fig: results}(c)), hazardous wind fields reach the site, prompting the model to initiate turbine shutdowns. In response, turbine blades are pitched to 90 degrees to reduce aerodynamic loads, nacelles are yawed to align with the incoming wind direction (approximately 218 degrees), and the turbine status transitions to \textit{Parked}.

This operational transition is formally captured using RDF triples, as illustrated in our event  domain ontology ($\mathcal{O}_{\text{EVENT}}$) (Figure \ref{fig:ontology}(b)). Taking Turbine~\#115 as an example, the ontology structure records the updated state via the triple:  
\textit{(:Turbine115 :hasTurbineStatus :Parked)}  
The RDF graph represents this turbine as a URI-linked entity, annotated with attributes such as pitch, yaw, wind speed, and drive type. The red arrow highlights the change in operational status, showcasing how simulation outputs are both machine-readable and semantically consistent with the OWL-defined schema. This confirms that turbine state changes are fully traceable within the semantic model and can be leveraged for automated control and monitoring. In Phase 3 (Figure\ref{fig: results}(d)), turbines safely recover and return to their operational state as wind conditions stabilize.

\section{Discussion and Conclusions}

Our study presents LSDTs, a framework that integrates LLMs with DT architectures to address the challenge of retrieving unstructured regulatory and technical information for infrastructure planning. By developing an LLM-driven ontology extraction pipeline, our approach transforms complex planning documents into structured semantic graphs, which enables DTs to support regulation-aware decision-making. Our pilot study using the Maryland offshore wind farm case demonstrates that LSDTs can automate the integration of diverse planning constraints and support efficient layout simulation in different scenarios.

Beyond static validation, our results demonstrate that integrating LLM-extracted knowledge with dynamic simulation and compliance reasoning enables DTs to adapt to evolving scenarios, such as extreme weather events. The hurricane simulation illustrates how the LSDTs workflow can support simulation of infrastructure resilience and generate automated in response to emerging hazards. This highlights the system’s capacity for adaptivity across both the planning and operational phases even in complex environments.

Looking ahead, the generalizable architecture of LLM-driven DTs systems can support broader application in domains where policies and regulations are originally unstructured but require structured extraction, such as transportation and urban development. Our framework demonstrates that combining LLMs with DTs can accelerate data-driven, regulation-compliant planning and decision-making. However, key challenges remain in ensuring the reliability of LLM outputs and adapting the system to diverse environments. Future work could refine automated rule formalization and engage stakeholders to ensure alignment with societal values and legal standards.

\section*{Ethical Statement}

The deployment of LSDTs for infrastructure planning presents both positive and negative societal implications. Our work provides key insights for more structured, efficient, regulation-compliant infrastructure design. By automating regulatory knowledge integration, LSDTs may streamline complex planning and reduce the efforts of manual compliance checking, particularly benefiting smaller organizations with limited technical resources. However, several limitations warrant careful consideration. Relying on LLMs to interpret regulations may introduce or amplify misinterpretations or omissions originating from training data or source documents. Errors in automated knowledge extraction or rule encoding could lead to flawed decision-making, with significant consequences in high-stakes environments. Moreover, the use of DTs and LLM-driven decision-making may also shift accountability from human planners to automated systems, raising concerns about transparency and responsibility. Therefore, system outputs must be subject to expert oversight, with safeguards in place to detect and correct errors. 

\section*{Acknowledgments}

This research was partially funded by the Department of Defense (DoD) Minerva Research Initiative, under the project ``Semantic Foundations and Formal Methods for Evolutionary System-of-System Architectures,'' with support from January to March 2025. We also acknowledge the valuable contributions of Juan Li, a software engineer at Google, who assisted with data cleaning and preprocessing, significantly enhancing the quality of the dataset used in this work.

\bibliography{aaai2026}

\end{document}